\documentclass[twocolumn,showpacs,amsmath,amssymb,pre]{revtex4}

\def\dbar{{\mathchar'26\mkern-12mu d}}

\usepackage[usenames,dvipsnames]{xcolor}
\usepackage{amsthm}
\usepackage{graphicx}
\usepackage{dcolumn}
\usepackage{bm}
\usepackage{graphicx}
\usepackage{epsfig}
\usepackage{color}
\usepackage{nicefrac}

\newcommand{\be}{\begin{equation}}
\newcommand{\ee}{\end{equation}}
\newcommand{\bea}{\begin{eqnarray}}
\newcommand{\ba}{\begin{array}}
\newcommand{\eea}{\end{eqnarray}}
\newcommand{\bes}{\begin{subequations}\bea}
\newcommand{\ees}{\eea\end{subequations}}
\newcommand{\ea}{\end{array}}

\providecommand{\diag}{\mathrm{diag}\,}

\newcommand{\bs}[1] {\boldsymbol{#1}}

\begin{document}

\newtheorem{theorem}{Theorem}


\title{Nonconvexity of the relative entropy for Markov dynamics: \\ A Fisher information approach} 

\author{Matteo Polettini} 
 \email{matteo.polettini@uni.lu}
\affiliation{Complex Systems and Statistical Mechanics, University of Luxembourg, Campus
Limpertsberg, 162a avenue de la Fa\"iencerie, L-1511 Luxembourg (G. D. Luxembourg)} 

\author{Massimiliano Esposito}
\affiliation{Complex Systems and Statistical Mechanics, University of Luxembourg, Campus
Limpertsberg, 162a avenue de la Fa\"iencerie, L-1511 Luxembourg (G. D. Luxembourg)}

\date{\today}

\begin{abstract}
We show via counterexamples that relative entropy between the solution of a Markovian master equation and the steady state is not a convex function of time. We thus let down a curtain on a possible formulation of a principle of thermodynamics regarding decrease of the nonadiabatic entropy production. However, we argue that a large separation of typical decay times is necessary for nonconvex solutions to occur, making concave transients extremely short-lived with respect to the main relaxation modes. We describe a general method based on the Fisher information matrix to discriminate between generators that do and don't admit nonconvex solutions. While initial conditions leading to concave transients are shown to be extremely fine-tuned,  by our method we are able to select nonconvex initial conditions that are arbitrarily close to the steady state. Convexity does occur when the system is close to satisfy detailed balance, or more generally when certain normality conditions of the decay modes are satisfied. Our results circumscribe the range of validity of a conjecture proposed by Maes {\it et al.} [Phys. Rev. Lett. {\bf 107}, 010601 (2011)] regarding monotonicity of the large deviation rate functional for the occupation probability (dynamical activity), showing that while the conjecture might still hold in the long time limit, the dynamical activity is not a Lyapunov function.
\end{abstract} 



\maketitle

\section{Introduction} 

The quest for general variational principles of thermodynamics and for arrows of time far from equilibrium leads researchers to sieve the behavior of several ensemble and path observables, in order to establish the stability of steady states, describe fluctuations out of them,  and characterize evolution towards them \cite{klein,glans,jiuli,maes1,maes2,maxent,plastino}. In the context of the probabilistic formulation of thermodynamics in terms of Markov processes  \cite{schnak,seifert,coarse}, blending aspects of information theory and thermodynamics, relative entropy with respect to the steady state naturally draws the inquirer's attention, having a threefold role: a dynamic one as a Lyapunov functional \cite{schnak}, a thermodynamic one as a nonadiabatic contribution to the entropy production \cite{esposito0,esposito1}, and a statistical one as a tool for parameter estimation \cite{kullback,fisher}. Along these lines, many may have conducted systematic research on the hypothesis that relative entropy is a convex function of time along the solution of a Markovian master equation, at least not too far from the steady state. Indeed, this is a tempting hypothesis, in that it would make for a new principle of thermodynamics, analogous to the principle of minimum entropy production \cite{jiuli,polettini}.

In this paper we will work with continuous-time, discrete-state space stationary Markov processes, described by a master equation whose solution $p(t)$ tends asymptotically to a unique steady state $p^\ast$. Along this solution, relative entropy with respect to the steady state is defined as
\be
H(t) =  \sum_i p_i(t) h_i(t), \label{eq:relent}
\ee
where we refer to
\be
h_i(t) = \ln \frac{p_i(t)}{p_i^\ast}
\ee
as the {\it relative self-information}.

Relative entropy is positive when $p(t) \neq p^\ast$, and it decreases monotonically to zero; hence it is a proper Lyapunov function \cite{schnak,vankampen}. For systems whose transition rates satisfy the condition of detailed balance, affording an equilibrium steady state with no net circulation of currents, relative entropy is convex when the system is sufficiently close to the steady state, i.e. in the linear regime. For this class of systems there exists an energy function and an environment temperature $T$, such that $dF = T dH $ is consistently identified as a free energy increment (setting Bolzmann's constant $k_B =1$). In this case the entropy production is a state function $\dbar S_i =- dH $, describing in many respects the system's thermodynamics. Monotonicity of the relative entropy corresponds to a positive entropy production rate  $\dot{S}_i \geq 0$, i.e. to the second law of thermodynamics. The entropy production rate vanishes at equilibrium, $\dot{S}_i = 0$, where no irreversible fluxes occur. Convexity of the relative entropy in the linear regime yields the stability criterion $\ddot{S}_i \leq 0$, which constitutes a version of the minimum entropy production rate principle \cite{jiuli}. Hence, the thermodynamics of systems relaxing to equilibrium states is fully encoded in the behavior of the relative entropy.

For autonomous nonequilibrium systems, whose generator does not depend explicitly on time, (minus) the time derivative of the relative entropy is not a fully satisfactory concept of entropy production rate, as one expects that nonequilibrium steady states should display a non-null steady flux of entropy towards the environment. It can still be interpreted as a nonadiabatic contribution $d S_{na} = - d H$ to the total entropy production $\dbar S_i = \dbar S_a + d S_{na}$, owing its name to the fact that, when the system is perturbed on time-scales that are longer than the spontaneous relaxation times of the system (adiabatic limit), this contribution approximately vanishes \cite{esposito1}. It has also been interpreted as a sort of nonequilibrium free energy for systems subject to nonequilibrium forces such as chemical potential gradients, in an isothermal environment \cite{ge1}. Along the lines of research developed by Schnakenberg \cite{schnak}, an adiabatic term $\dbar S_a$ is added to the nonadiabatic one. It accounts for a flux of entropy from the system to the environment, due to departure from the condition of detailed balance. The adiabatic/nonadiabatic splitting of Schnakenberg's entropy production is particularly useful to characterize the second law when non-autonomous (nonstationary) Markovian evolution is considered \cite{verley}, in which case one shall also account for a work term \cite{espoepl}.  

In general, the total entropy production rate is not a state function, reflecting the well-known fact that irreversibility along nonequilibrium processes is characterized by inexact differentials that do not integrate to zero along closed paths, a notable example being Clausius's formulation of the second law $\oint \delta Q /T \geq 0$ (see Ref. \cite{polettini3} for a discussion in the context of Schnakenberg's theory). For this reason we distinguished between the exact and the inexact differentials $d$ and $\dbar$.  Moreover, while positive in accordance with the second law of thermodynamics, the total entropy production rate presents no apparent regularity in its time evolution. In particular it does not approach its steady value monotonously.

Then, convexity of the relative entropy, at least in the linear regime, is an intriguing  hypothesis, in that it would make for a minimum principle of a nonequilibrium state function, amending  the unpredictable behavior of the entropy production rate. The principle would state that  the nonadiabatic rate of entropy production decreases monotonously to zero, regardless of the concomitant spontaneous arrangement of heat fluxes, matter fluxes, charge currents etc.

\subsection{Results and plan of the paper}

Is relative entropy convex? We answer in the negative, providing in Sec.\ref{counter} a counterexample for a simple three-state system with real spectrum of the generator. In Sec.\ref{real} we show that convexity violation can occur with initial conditions picked arbitrarily close to the nonequilibrium steady state. By close we mean that the second order term in the expansion of the relative entropy captures the full dynamical behavior. However, we argue in Sec.\ref{separation} that convexity violation is rare and short-lived. It requires a wide separation of time scales and a very fine tuning of the initial conditions. Moreover, an extensive numerical search did not allow us to find counterexamples for three-state systems with complex spectrum, which might indicate that convexity is even more robust when some eigenmodes have an oscillatory character.

Throughout the paper we employ a method based on the Fisher covariance matrix to characterize generators that violate convexity. Systems with real spectrum are described in Sec.\ref{real}. We discuss the special case of equilibrium systems in Sec.\ref{equilibrium}.  For sake of completeness, the general theory for systems with complex spectrum is analyzed in Appendix \ref{complex}. Convexity still holds near the steady state for a special class of ``normal'' systems, including detailed balanced systems, as Maes {\it et al.} discussed \cite{maes1}; we recast this result in our formalism.

A connection between the second time derivative of the relative entropy and the first time derivative of the dynamical activity near the steady state is established  in Sec.\ref{dyn}. This allows us to discuss the range of validity of a conjecture by Maes {\it et al.} regarding the monotonicity of the dynamical activity \cite{maes1,maes2}. 

If present at all, nonconvex transients of the relative entropy and nonmonotone transients of the dynamical activity prelude to a final regime dominated by the mode with slowest decay rate. This regime is trivially convex for systems with real spectrum, while in the complex case one encounters interesting complications, leading to conditions on the real and imaginary parts of complex conjugate eigenvalues. We briefly pursue this discussion in Sec.\ref{longtime}.

Finally, in Sec.\ref{fisher} we briefly discuss the rationale behind the use of the Fisher information measure, before drawing conclusions.

\subsection{\label{counter}Counterexample}

\begin{figure}
  \centering
 \includegraphics[width=240pt]{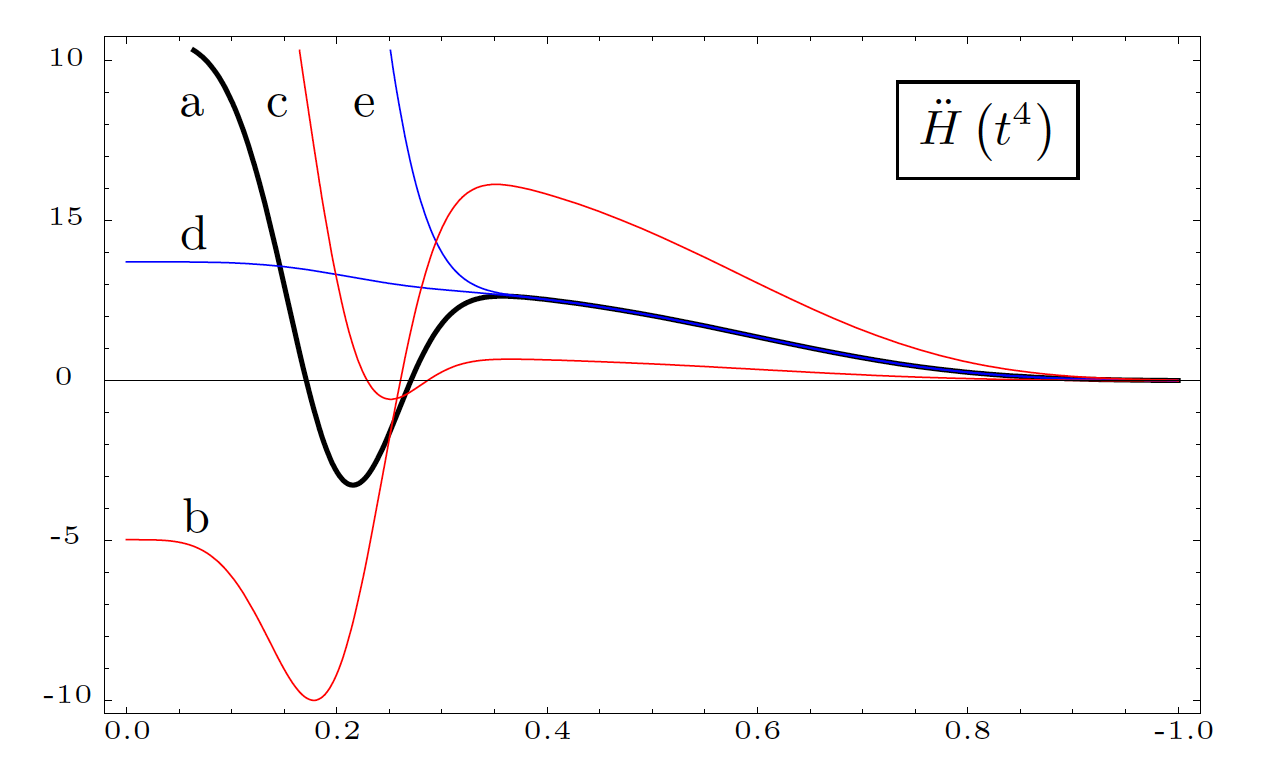}
  \caption{ \label{immagine} Second time derivative of the relative entropy as a function of $t^4$, with different initial conditions: (a) As in our counterexample, $p^{(a)}(0) = p$; (b,c) Perturbed along the slow mode, with $p^{(b)}(0) = p + 0.1 \, q_-$, and $p^{(c)}(0) = p - 0.1 \, q_-$; (d,e) Perturbed along the fast mode, with  $p^{(d)}(0) = p - 0.001 \, q_+$ and $p^{(e)} = p -0.005 \, q_+$.}
\end{figure}

Consider the continuous-time Markovian generator
\be
W =  \left(
\ba{ccc}
-401 & 1 & 1 \\
400 & -2 & 1 \\
1 & 1 & -2
\ea
\right), \label{eq:gen}
\ee
with steady state
\be
p^\ast = (3,801,402)/1206.
\ee
Notice that the system is strongly unbalanced, with one overwhelmingly large rate. As a consequence, one state is almost neglected, its occupancy probability falling rapidly to a value near zero. We choose as initial density
\be
p = (0.002, 0.464, 0.534).
\ee
We propagate $ {p}$ in time via $ {p}(t) = \exp (tW)  {p}$, and evaluate relative entropy with respect to the steady state. The plot of $\ddot{H}(t)$ in Fig.\ref{immagine} (bolder line) clearly becomes negative for a short transient time.

The above generator has a real spectrum, with eigenvalue zero relative to the steady state and two negative eigenvalues $\lambda_+ = -402$ and $\lambda_- =-3$ determining respectively fast and slow exponential decays. Hence, the system displays a large separation between typical decay times. The corresponding eigenvectors are:
\be
 q_+ \approx (-1,1,0), \qquad  q_- \approx (0,-1,1).
\ee
If we perturb the initial condition along the mode with the slower decay rate $q_-$, even large perturbations do not suffice to restore convexity (see curves (b) and (c) in Fig.\ref{immagine}). However, if we perturb the initial condition along the mode with the faster decay rate $q_+$, even  slight perturbations do (Fig.\ref{immagine}, curves (d) and (e)). Thus a very precise fine-tuning on the initial conditions must be attained to generate a counterexample. Moreover, since the dynamics damps the fastest mode first, the concave regime is extremely short-lived, with a typical survival time of order $\tau \sim -\lambda_+^{-1}$. Finally, the large separation of time scales implies that mapping the initial state back in time with $\exp (-tW)$ leads very soon to to nonphysical solutions (negative probabilities), as the fastest eigenmode would now dominate. Reversing the argument, ``typical'' dynamics will not pass by state $p$, which has to be specifically selected.

\section{Theory and results: Real spectrum}

\subsection{\label{real}Conditions for convexity violations}

We describe in this section a general algebraic procedure that allows to discriminate between generators that do or don't admit initial conditions violating convexity.

Let us consider a Markovian continuous-time evolution on $n$ states, with irreducible rates $w_{ij}$ for jumps from state $j$ to $i$ admitting a unique steady state $p^\ast$. It is known that the $n-1$ non-null eigenvalues $\lambda_a$ of $W$ have a negative real part with units of an inverse time $- 1/\tau_a$, characterizing relaxation. We consider in this section generators with real non-degenerate spectrum, which afford a complete set of independent eigenvectors. We will discuss defective generators, affording a nondiagonal Jordan normal form, in Sec.\ref{defective}. The eigenvalue equations read
\be
W  q^a = \lambda_a  q^a, \qquad W  p^\ast = 0.  \label{eq:eigeneqs}
\ee
Diagonalizing the propagator $U(t) = \exp (tW)$, we obtain for the time-evolved distribution 
\be
p(t) ~=~  p^\ast + \sum_a e^{\lambda_a t} c_a  q^a, 
\ee
where $\bs{c} =(c_1,\ldots,c_{n-1})$ is a real vector, specifying the initial state of the system. Let $c_a(t)= c_a \exp \lambda_a t$.  We expand the relative entropy to second order, obtaining
\be
H(t) \approx \sum_{a,b} g^{ab} c_a(t) c_b(t), \label{eq:relent2}
\ee
where
\be
g^{ab}  = \frac{1}{2}\sum_i  \frac{q^{a}_i q^{b}_i}{p_i ^{\,\ast}} =  \frac{1}{2} \langle q^a , q^b \rangle.  \label{eq:fisher} 
\ee
The right-hand side defines a scalar product $\langle  \,\cdot \,,\,\cdot \, \rangle$.
Properties of the matrix $g^{ab}$, expecially regarding equilibrium systems, are well known \cite[Sec. 5.7.]{vankampen}. 
Here it will be called {\it Fisher matrix} for reasons that are rooted in estimation theory, and that will be explained in Sec.\ref{fisher}. It is a Gramian matrix, i.e. its entries are obtained as scalar products among vectors. When vectors $q^a$ are independent, as under our assumptions, Gramian matrices are positive definite  \cite{horn}. The Fisher matrix can then be seen as a realization in local coordinates of a metric on the space of statistical states; some applications to nonequilibrium decay modes have been discussed by one of the authors in Ref. \cite{polettini2}.

We introduce the negative-definite diagonal matrix of eigenvalues
\be
\Lambda = \mathrm{diag\,} \left\{\lambda_1, \ldots, \lambda_{n-1}\right\}.
\ee
Let $\bs{c}(t) = e^{t \Lambda} \, \bs{c}$. Taking twice the time derivative of the relative entropy, to second order, we obtain
\be
\ddot{H}(t) = \bs{c}(t)^T \stackrel{\;K}{\overbrace{ \Big( 2 \Lambda G \Lambda +  \Lambda^2 G + G\Lambda^2 \Big)}} \bs{c}(t).\ee
The overbrace is used to define a bilinear symmetric form $K$, whose first contribution $2 \Lambda G \Lambda $ is positive definite. However, $K$ itself might admit at least one eigenvector $\bs{k}$ relative to a negative eigenvalue. When this is the case, the choice of initial conditions  $\bs{c} \propto \bs{k}$ yields an initially negative second time-derivative of the relative entropy. Moreover, since the length of $\bs{c}$ can be made small at will,  we can select initial states that are arbitrarily close to the steady state, still displaying violation of convexity, and fulfilling the second-order approximation to any degree of accuracy. Since $K$ is known to be positive for particular systems, by continuity a negative eigenvalue of $K$ can only occur if there exist generators such that
\be
\det K = 0. \label{eq:det}
\ee
Notice that  $K$ is built out of the eigenvalues and eigenvectors of the generator, which are expressed in terms of transition rates. Hence Eq.(\ref{eq:det}) identifies an algebraic set within the set of allowed rates.

To recapitulate, the search for nonconvex generators is reduced to an algebraic polynomial equation, whose difficulty can be tuned at will by suitably parametrizing transition rates. Once a generator with at least one negative eigenvalue of $K$ is found, one can solve the eigenvalue/eigenvector problem and find initial conditions that violate convexity. We report that this procedure greatly reduced the computational complexity of the problem: Rather than randomly searching for a generator \textit{and} an initial state, we only looked for a suitable generator by a simple algebraic procedure; nonconvex initial conditions follow.

\subsection{\label{equilibrium}Time reversal close to equilibrium}

In this paragraph we show that the relative entropy of close-to equilibrium systems obeys convexity, and how properties of the Fisher matrix are related to time-reversal symmetry. This analysis will be extended to complex spectrum case in Appendix \ref{normal}. 

Given that dissipative dynamics have a preferred time direction, the time-reversal generator  \cite[p.47]{denmar} is what comes closest to reverting the direction of time of a Markov process, by inverting certain nonequilibrium characteristics (e.g. steady currents) while preserving others (e.g waiting times). It has been considered by various authors in relation to fluctuation theorems \cite{crooks1,ccj,esposito2}, to prove convexity for normal systems \cite{maes1}, to discuss spectral properties of Markov processes \cite{andrieux}, and to identify a supersymmetry in Markovian dynamics \cite{sinitsyn}.

The definition of time-reversal is more intuitive in the adjoint picture. Consider the steady state correlation of two functions  $\sum_i p_i^\ast f_i g_i$ at time $t=0$, and  displace $f$ with the adjoint generator for a small time,  $f(d t) = (1 +  W^{T}  d t) f  $. We ask which generator should be employed to evolve $\overline{g}(dt) = (1 + \overline{ W}^T dt) g$ in such a way that the following time-reversal identity holds
\be
\sum_i p^\ast_i f_i(dt) g_i(0) = 
\sum_i p^\ast_i f_i(0) \overline{g}_i(d t). \nonumber
\ee
The latter equation defines the adjoint of the time-reversed generator  $\overline{ W}^T$. Introducing the diagonal matrix
\be
P = \diag \{p^{\ast}_1,\ldots,p^{\ast}_{n-1}\} ,
\ee
an explicit calculation shows that the time-reversal generator is given by
\be \overline{ W} =  P\,  W^T P^{-1}.  \label{eq:timereversal} \ee
Some of its properties are: The transformation is involutive; The reversed dynamics affords the same steady state; Steady currents and affinities change sign; Exit probabilities out of states are unchanged; The spectra of $\overline{W}$ and $W$ coincide.

Equilibrium generators are those for which the time-reversed generator coincides with the original generator, $W = \overline{W}$. In practice, this translates into the condition of detailed balance $w_{ij} p_j^\ast = w_{ji} p_i^\ast$. We can further characterize equilibrium generators in terms of the Fisher matrix as follows. We define
\be
V = P^{-1/2} W P^{1/2} . \label{eq:sim}
\ee
Equation (\ref{eq:sim}) is a similarity transformation, hence the spectra of $W$ and $V$ coincide, and eigenvectors are mapped into eigenvectors. Performing an analogous transformation on the reversed generator we obtain
\be
V^T = P^{-1/2} \bar{W} P^{1/2} = P^{1/2}  W^T   P^{-1/2}.
\ee
Hence the condition of detailed balance translates into $V$ being symmetric. By the spectral theorem it follows that its spectrum is real (equilibrium systems do not admit complex eigenvalues), and it affords a complete set of orthonormal eigenvectors $v^a$. Letting $v^0 = \sqrt{p^\ast}$ be the null eigenvector of $V$, all other eigenvectors $v^a$ can be normalized so to have
\be
2g^{ab} = \sum_i v^a_i v^b_i = \delta^{ab}.  \label{eq:euclidean}
\ee
On the left-hand side one can recognize the Fisher matrix, by transforming back to the eigenvectors of $W$, given by $q^a = P^{-1/2} v^a$. This transformation maps the euclidean scalar product in the above equation into the scalar product $\langle \, \cdot \,, \, \cdot \,\rangle$. Given that we followed a chain of necessary and sufficient facts, it is then proven that the Fisher matrix $G$ is diagonal if and only if the generator $W$ satisfies detailed balance. In this case we have
\be
K = 4 \Lambda^2
\ee
which is obviously positive definite. Hence convexity holds for equilibrium systems, in the linear regime. We don't know whether a violation of convexity could occur out of the linear regime, where higher-order contributions from the logarithm in the expression for the relative entropy might come into play. By continuity, nearly equilibrium systems also satisfy convexity.

\subsection{\label{separation}Time-scale separation}

The counterexample provided in Sec.\ref{counter} is characterized by a wide separation of typical decay times. It is an interesting question whether time-scale separation is necessary for violating convexity. It certainly is not sufficient, as it is well known that the spectrum alone does not characterize the nonequilibrium character \cite{andrieux}. For example the generator
\be
W = \left(
\ba{ccc}
-202 & 201 & 1 \\
201 & -202 & 1 \\
1 & 1 & -2
\ea
\right),
\ee
has decay times $1/3$ and $1/403$, but it satisfies detailed balance, hence it does not violate convexity.

In the rest of this section we argue that a large time-scale separation might be necessary. We first hint at a general argument in favor of this conjecture, and then discuss the complications arising with nearly defective generators in the next section.

Two consequences of time-scale separation are that concavity is extremely short-lived and that it has a short past. In fact, selecting initial conditions $\bs{c}= \bs{k}$ along a negative eigenvector of $K$, and perturbing them for a short time $\tau$, the time evolved coefficients $\bs{c}(\tau) \approx (1 + \Lambda \tau) \bs{k}$ skew $\bs{k}$ along the fastest mode, with typical time for restoring convexity given by the smaller decay time, $\tau = - \sup_a \lambda_a^{-1}$. For the same reason, mapping back in time with $\exp(-tW)$ leads soon to negative, nonphysical probabilities. Hence, nonconcave states are hardly encountered by ``typical dynamics''. 

Let us suppose that decay rates are not widely separated, i.e. that there exists some average value $\lambda$ within the spectrum such that
\be
\epsilon_{i} = \frac{\lambda_i - \lambda}{\lambda}
\ee
are all small.  Defining the matrix
\be
\varepsilon = \mathrm{diag}\, \{ \epsilon_1, \ldots, \epsilon_{n-1} \},
\ee
such that $\Lambda = \lambda (I - \varepsilon)$, with $I$ the $(n-1)$-dimensional unit matrix, and evaluating
\bes
\Lambda G \Lambda & = & \lambda^2 \left( G +  \varepsilon G + G \varepsilon + \varepsilon G \varepsilon \right) \\ 
\Lambda^{\!2} G & = & \lambda^2 \left( G + 2 \varepsilon G + \varepsilon^2 G \right) \\
G \Lambda^{\!2} & = & \lambda^2 \left( G + 2 G \varepsilon + G \varepsilon^2 \right),
\ees
we find that
\be
K  = 4\Lambda G \Lambda + \lambda^2 [\varepsilon, [\varepsilon, G]]. \label{eq:second}
\ee
Here $[\cdot,\cdot]$ is the commutator. Equation (\ref{eq:second}) states that corrections to the positive definite contribution $4\Lambda G \Lambda$  are second-order in the eigenvalue spacings. If $ {w}^T \Lambda G\Lambda  {w}$ is finite for all vectors $ {w}$ of finite norm, there are no possible initial conditions that will lead to a violation of convexity.

\subsection{\label{defective}Time-scale separation: Defective generators}

\begin{figure}
  \centering
    \includegraphics[width=240pt]{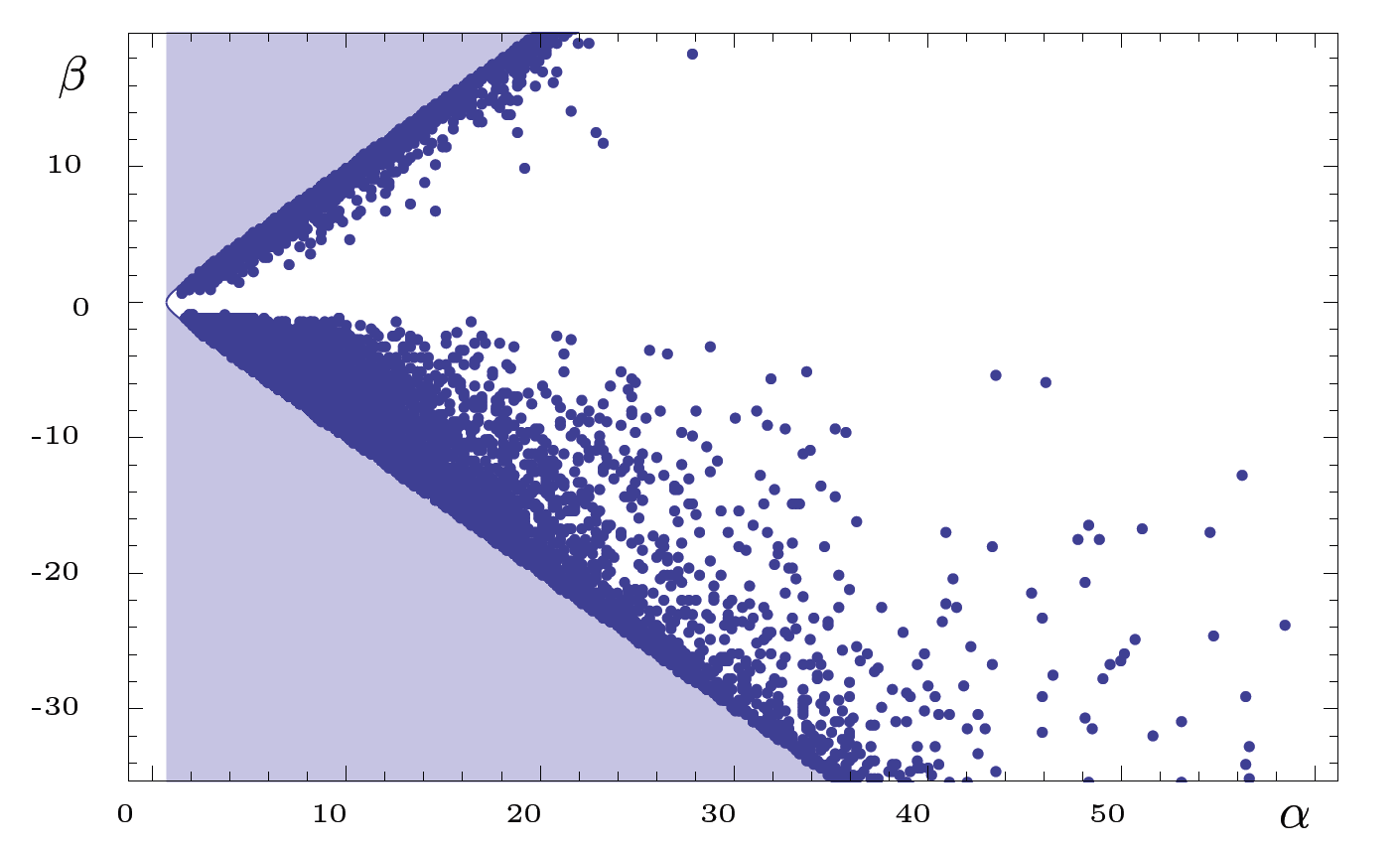}
  \caption{\label{fig2}20\,000 randomly generated values of $\alpha,\beta$. The shaded region corresponds to $\alpha^2 - \beta^2 \leq 1/2$.}
\end{figure}

However, it can be the case that when eigenvalues are made closer to each other, eigenvectors of the generator also tend to overlap, in a limit where $W$ becomes defective, as it lacks a complete set of eigenvectors relative to degenerate eigenvalues.  As one of the authors analyzed in Ref.\cite{polettini2}, when $W$ is nearly defective, $G$ is nearly degenerate
and it affords a nearly null eigenvector $w_0$. For example for three-state systems, one would have
\be
G \sim \left( \ba{cc} 1 & 1+ O(\epsilon^2) \\ 1 + O(\epsilon^2) & 1 \ea \right) \label{eq:gdeg}
\ee
and $w_0 = (1,-1)$. Then choosing $w = \Lambda^{-1} w_0$ makes ${w}^T \Lambda G\Lambda  {w}$ small of order $\epsilon^2$ as well. Notice that, among matrices with degenerate spectrum, matrices that afford a basis of eigenvectors are a set of zero measure with respect to defective matrices \footnote{This is because blocks in their Jordan form have lower algebraic than geometric multiplicity for all values of the non-null off-diagonal terms.}, so the latter are quite crucial for our argumentation.

In this section, for three-state systems, we bring computational evidence that slightly departing from a defective generator, convexity still holds for any initial conditions, so that one might conclude that nonconvexity and time scale separation go by hand. A general proof seems to be elusive.

Consider a generic three-state system with real spectrum, with two real eigenvalues $-\lambda_{\pm}$ relative to eigenvectors $ {q}_{\pm}$. Matrix  $K$ is given by
\be
K = \left(\ba{cc} 4 \lambda_+^2 \langle  {q}_+, {q}_+ \rangle & (\lambda_+ + \lambda_-)^2\langle  {q}_+, {q}_- \rangle \\  (\lambda_+ + \lambda_-)^2 \langle  {q}_+, {q}_- \rangle  & 4 \lambda_-^2 \langle  {q}_-, {q}_- \rangle\ea \right), 
\ee
and the determinant condition for convexity reads
\be
\frac{4\lambda_+ \lambda_-}{(\lambda_+ + \lambda_-)^2} > \frac{|\langle  {q}_+,  {q}_- \rangle |}{||\, {q}_+ || \cdot ||\, {q}_-||} = \cos \varphi, \label{eq:dise}
\ee
where the norm is calculated with respect to the scalar product $\langle\cdot,\cdot\rangle$. We recognize in the right-hand side the cosine of the angle $\varphi$ between the two vectors, which vanishes when eigenmodes are orthogonal, i.e. for equilibrium systems. At the opposite extremum, $\cos \varphi$ reaches value $1$ when eigenmodes are collinear; this only occurs when the system is defective, as it lacks a set of independent eigenvectors, in which case the two eigenvalues are identical, and also the left-hand side attains value $1$. In the vicinity of a defective generator, with slightly-spaced eigenvalues and eigenvectors
\be
\lambda_{\pm} =\lambda(1 \pm \epsilon), \qquad  {q}_{\pm} =  {x} \pm \epsilon  {y},
\ee
we have $4\lambda_+ \lambda_-/(\lambda_+ + \lambda_-)^2 = 1 -  \epsilon^2$ and
\be
\cos \varphi =  1 - \frac{2 \epsilon^2}{||\, {x}\,||^4}\left(||\, {x}\,||^2 ||\, {y}\,||^2 - \langle {x}, {y} \rangle^2 \right).
\ee
Introducing the two parameters $\alpha = ||\, {y}\,||/||\, {x}\,||$ and $\beta = \langle {x}, {y} \rangle/ ||\, {x}\,||^2$, disequality (\ref{eq:dise}) becomes
\be
\alpha^2 - \beta^2 > 1/2. \label{eq:disab}
\ee
In Appendix \ref{app2} we give further details on how to express $x$ and $y$ (hence $\alpha$ and $\beta$) in terms of the transition rates of a nearly defective generator. In Fig.\ref{fig2} we plotted $20\,000$ randomly generated values of $\alpha,\beta$ for defective generators, finding that none of them violates disequality (\ref{eq:disab}). This suggests that, at least for three-state systems, a large separation of time scales is necessary.

\section{Additional topics}

\subsection{\label{dyn}Nonmonotonicity of the dynamical activity}

\begin{figure}
  \centering
   \includegraphics[width=240pt]{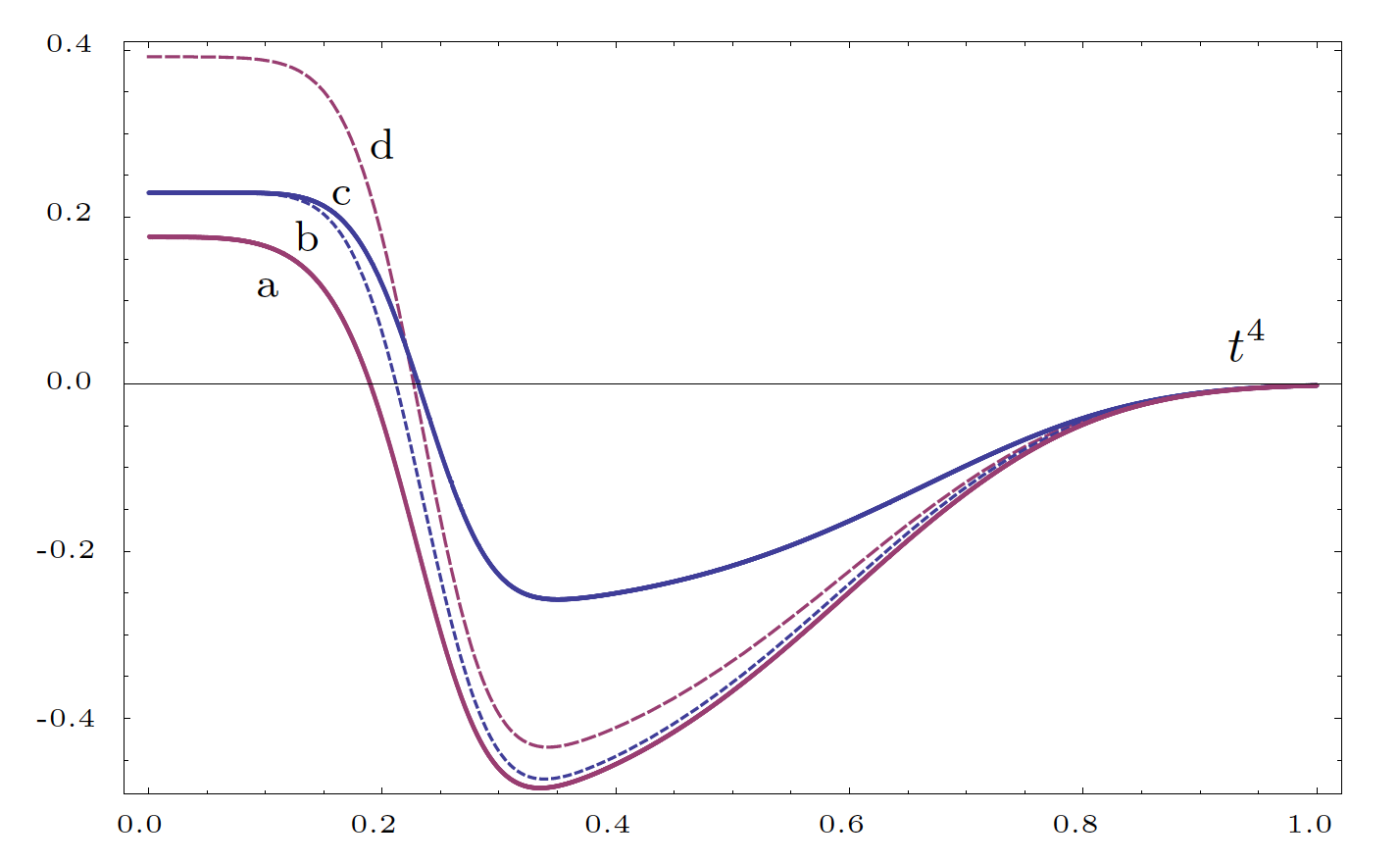}
  \caption{ \label{maes}In solid lines, plotted as functions of $t^4$, the first derivative of (a) the dynamical activity as a function of $\mu(t)$, and (c) its second-order approximation. In dashed lines, (d) one-half minus the second time derivative the relative entropy as a function of $p(t)$, and (b) its second order approximation.}
\end{figure}

In the light of the above results, in this section we will present a discussion of the work by Maes, Neto\v cn\'y and Wynants  \cite{maes1,maes2}, to which we refer for the details. 

Consider a stochastic trajectory $\omega$ up to time $t$, and let $\mu^\omega_i(t)$ be the fraction of time spent on site $i$ along the trajectory. The question is, how typical is a set of values $\mu^\omega(t)\equiv \mu(t)$? The answer is found in the framework of large deviation theory in terms of the Donsker-Varadhan rate functional or {\it dynamical activity} $D[\mu(t)]$, which describes fluctuations out of the most probable distribution  $p^\ast$, for which it vanishes. Here, time plays the role of the extensive parameter; as $t \to \infty$, the steady state is exponentially favored. Instead, if $N$  trajectories are independently sampled at fixed time $t$, the rate functional for the probability of being at a site is given by relative entropy, with $N$ the extensive parameter. Therefore, relative entropy is a static rate functional while the dynamical activity is a dynamic one. The latter captures the Markovian nature of the process, while the former regards independent realizations.

All this given, the static nature of relative entropy as a large deviation functional does not prevent it from being monotonically decreasing, when evaluated along the solution of a Markov process \cite{schnak}.  Does the Donsker-Varadhan functional monotonically decrease as well? Maes {\it et al.} showed that it doesn't when the initial state is picked far from the steady state, but that monotonic behavior is restored in the long time limit. They discussed a few examples supporting their case, and proved that normal systems (see Appendix \ref{normal}) display a longstanding monotonic behavior. The remaining question is whether monotonicity might occur with initial states picked along any superposition of modes, arbitrarily close to the steady state. Leading back to our previous counterexample, we will show that this is not the case. More specifically, our counterexample shows that one of the hypothesis for Theorem III.1 in Ref. \cite{maes1} is not generally satisfied. 

Let $\mu(0)$ be the initial state of the system, $\mu(t) = \exp (Wt) \mu(0)$ be its time-evolved, and let us consider a time-dependent transformation of the transition rates
\be
w^{u(t)}_{ij} = w_{ij} \, e^{[u_i(t) - u_j(t)]/2}, \label{eq:modrates}
\ee
such that the generator $W^{v(t)}$ simulates steadiness at a frozen time $t$, that is, it affords $\mu(t)$ as its steady state,
\be
W^{u(t)}  \mu(t)  = 0.  \label{eq:transss}
\ee
It can be proven that there exists a unique choice of $u(t)$, up to a ground potential, such that the above equation holds. Maes {\it et al.} proved that the Donsker-Varadhan functional is given by
\be
D[\mu(t)] = \sum_{i,j} \left[ w_{ij} - w_{ij}^{u(t)} \right] \mu_j(t).
\ee
It affords a simple interpretation as the difference between the average escape rate of the actual dynamics and that of the time-frozen steady dynamics at time $t$. According to Eq.(10) in Ref.\cite{maes2}, when the state of the system is sufficiently close to the steady state, one has
\be
\dot{D}[\mu]  
= -\frac{1}{2} \left[  \Big(PW^T u, W^T u \Big) + \Big(Pu,{W^T}^2 u \Big) \right],
\ee
where we remind that $P$ is the matrix having the steady probability entries along its diagonal. Letting $u = P^{-1} p $, after some manipulations we obtain
\be
\dot{D}[\mu] = - \frac{1}{2}\left\langle (\overline{W} + W )p, \overline{W} p\right\rangle.
\ee
Similarly, we can express the second time derivative of the relative entropy to second order as
\be
\ddot{H}[\,p\,] = \left\langle (\overline{W} + W )p, W p\right\rangle.
\ee
We notice an analogy by interchanging the generator and its time reversed. Therefore, to second order
\be
\frac{d}{dt} D[\exp(tW) \mu(0)] \Big|_0 = - \frac{1}{2} \frac{d^2}{dt^2} H[\exp(t\overline{W}) p(0)] \Big|_0. \label{eq:comparison}
\ee
The respective initial conditions are connected by
\be
W^{P^{-1} p(0)} \mu(0) = 0. 
\ee

In Fig.\ref{maes} we represent violation of monotonicity of the dynamical activity, using as time-reversed generator the one already employed in Sec.\ref{counter}, namely  Eq.(\ref{eq:gen}). The initial state was chosen to have a relative entropy  2.5\% (in base $3$ \footnote{The choice of base $n=3$ is a normalization suggested by the fact that the relative entropy of the certain distribution $\delta_{i,1}$ with respect to the uniform distribution $1/n$ is $\ln n$.}). It is possible to reduce this measure of distance to smaller and smaller values, making all curves in the picture closer and closer. Notice that the correspondence between the dynamical activity and the relative entropy (to second order) only holds at the initial time. Later, the two differ for two reasons: Their time behavior is due to different dynamics; they are not the same functional of the probability distribution.

We point out that Lyapunov's second theorem for stability assumes that there exists a function whose first derivative is negative in some neighborhood of a candidate fixed point. Using our procedure, we were able to show that the dynamical activity and the first time derivative of the relative entropy do not satisfy this requisite.  We note however that  this fact is quite irrelevant, since  the stability of steady states  for irreducible Markov processes is  well-established.

\subsection{\label{longtime}Long-time behavior}

 Since  our results show that nonconvex transients of the relative entropy and nonmonotone transients of the dynamical activity can occur arbitrarily close to the steady state, a natural question is  whether a convex/monotone behavior is always restored in the long time limit when the dynamics is dominated by the mode with slowest decay rate. While this is a trivial fact for systems with real spectrum, for systems with complex spectrum it only holds when certain algebraic relations between the real and the imaginary parts of the eigenvalues are satisfied.

In the real spectrum case, let $\lambda_1$ be the largest eigenvalue that affords a nonnull coefficient $c_1$ in the expansion of the initial state. We assume $\lambda_1$ to be nondegenerate. Then at large times
$p(t) \sim p^\ast + c_1 \, e^{t \lambda_1} q^1$, and
\be
H(t) \sim g_{11} c_1^2 \, e^{2 t \lambda_1},
\ee
which is obviously convex.

The case of systems with complex spectrum is discussed at length in Appendix \ref{complex}. In general, the relative entropy can be written as a quadratic form in terms of a Fisher matrix that contains the information about the superposition of the real and imaginary parts of the complex eigenmodes. Let us only report that, by letting $\lambda_1 = -\tau_1^{-1} + i \omega_1$ and $\lambda_1^\ast$ be the complex conjugate eigenvalues with the largest real part affording nonnull coefficients $c_1,c_1^\ast$ in the expression for the initial state, $p = {p}^{\,\ast} +  c_1   {q}^{1}  + c^{\,\ast}_1   {q}^{1 {\,\ast}}$, convexity in the long time limit implies the following relationship between real and imaginary  parts of the eigenvalues:
\be
\left[\frac{1}{1 + (\omega_1\tau_1)^2}\right]^2 \geq 1 - \frac{4\, \mathrm{det}\, G_{1}}{\left(\mathrm{tr}\,G_{1}\right)^2}.
\ee
$G_{1}$ is the $2\times 2$ matrix having as entries the superpositions between real and complex parts of the relevant eigenvector, i.e. $\tilde{g}^{\, r_1i_1} = \langle \Re {q}^{1}, \Im {q}^{1} \rangle / 2$, and so on. Letting $g_+,g_-$ be the positive eigenvalues of $G_1$ with $g_+ \geq g_-$, the above condition translates into
\be
(\tau_1 \omega_1)^2 \leq \frac{2 g_-}{g_- - g_+}.
\ee
In particular, the period of oscillation is (variably) bounded from below by its corresponding relaxation time. If convexity is to hold in the long time limit, oscillations cannot be too fast with respect to their  typical exponential decay time. The upper frequency bound depends on how the real and complex part of the decay mode overlap.

\subsection{\label{fisher} On the information geometry of relative entropy and eigenmode estimation}

Before coming to conclusions, in this section we briefly linger on the interpretation of the Fisher matrix.

Let $\hat{x}$ be a random variable taking values $i$ with probability distribution $p_i(\vartheta)$ conditional on an unknown parameter $\vartheta$ whose value one might want to estimate. Fisher's information is defined as \cite{fisher}
\be
G(\vartheta) = \left \langle  \left( \frac{\partial p(\vartheta)}{\partial \vartheta} \right)^2 \right\rangle_{p(\vartheta)}.
\ee
It measures how much information the random variable retains about the parameter. The derivative with respect to $\vartheta$ grants that $G(\vartheta)$ detects the sensibility of the probability to a parameter variation. For example, if the probability distribution does not depend on the parameter at all, it vanishes. An important result concerning Fisher's information is that it sets a bound to the accuracy of an estimation of the parameter $\vartheta$ expressed by the Cram\'er-Rao inequality $G(\vartheta) \mathrm{Var} (\hat{\vartheta}) \geq 1$, where $\hat{\vartheta}$ is a so-called unbiased estimator of parameter $\vartheta$. For example, in the case where the probability does not depend on $\vartheta$, the variance of the estimator is infinite. This sort of indeterminacy relations have been put in contact with quantum \cite{luo} and statistical \cite{statun} uncertainty; see Ref.\cite{falcioni} for an application to temperature estimation.

It has also long been known \cite{kullback} that Fisher's information is twice the relative entropy (Kullback-Liebler divergence) $H\{\,\cdot \, \Arrowvert \, \cdot \,\}$ between two nearby probability distributions, to second order:
\be
H \left\{p(\vartheta + d \vartheta) \big\Arrowvert p(\vartheta)\right\} \approx \frac{1}{2} g(\vartheta)d\vartheta^2 .
\ee
While this point of view slightly hinders the statistical relevance for parameter estimation, it provides a clear geometrical picture since the relative entropy locally defines a metric on submanifolds of the space of statistical states. This metric is called the Fisher-Rao metric \cite{gibilisco}. Generalizing to several estimation parameters, we can express the Fisher-Rao metric in local coordinates $\bs{\vartheta} = (\vartheta_a)_a$ as  \footnote{For reasons of internal consistency we inverted the convention on index positioning with respect to standard literature on physical applications of differential geometry.}
\be
H \left\{ p(\bs{\vartheta} + d\bs{\vartheta})  \big\Arrowvert p(\bs{\vartheta} ) \right\} \approx \frac{1}{2} g^{ab}(\vartheta) d\vartheta_a d\vartheta_b.
\ee
The Fisher matrix then arises as one possible representation of the metric in a set preferred coordinates, which are dictated by the problem at hand. For example, in applications to equilibrium statistical mechanics the Fisher matrix takes the form of a covariance matrix, coordinates being the intensive variables conjugate to the physical extensive observables in the equilibrium measure (temperature, pressure, chemical potential, interaction constants etc.) \cite{equi}. Using the square roots of the entries of the probability density $\psi_i = \sqrt{p_i}$ as coordinates, one obtains the real part of the Fubini-Study metric for quantum states \cite{fubini}. Studies on the quantum Fisher information have also been proposed \cite{brody,ercolessi}. Beyond equilibrium, recent works \cite{crooks2} focus on how geodesic transport can represent classes of nonequilibrium transformations. Far from equilibrium, the Fisher information has been employed to characterize the arrow of time \cite{plastino}.

In our work, the Fisher matrix is obtained by parametrizing the probability distribution in the vicinity of the steady state with a vector of variables $\bs{\vartheta} = \bs{c}(t)$, at fixed time. Expressing the probability increment along a small displacement from the steady state as $dp = \partial^a p(\bs{\vartheta}) d\vartheta_a$, we can interpret the decay modes $q^a =  \partial^a p$ as tangent vectors at $p^\ast$ \cite{polettini2}. In this guise, the Fisher matrix tells how the probability density depends on eigenmodes, and in particular how eigenmodes are correlated. More specifically, perturbing the steady state in the $a$-th direction, and defining the relative  self-information carried by the $a$-th mode as
\be
h^a_i = \log \frac{p^\ast_i + \epsilon q^a_i}{p^\ast_i}  \approx \epsilon q^a_i/p^\ast_i, 
\ee
we can interpret the Fisher matrix $g_{ab}=  \epsilon^{-2}   \langle h^a h^b \rangle$ as the correlation matrix between single-mode self-informations, which are uncorrelated for normal systems, including equilibrium systems.

Within our framework, one can interpret the  Cram\'er-Rao inequality as a limit to the precision with which one can estimate the weight of a mode at some time, hence, more interestingly, as a limit on our ability to trace back the initial conditions. In this respect, eigenvalues play a role analogous to Lyapunov's exponents in dynamical systems. Suppose we want to estimate the permanence of eigenmodes in a state at some given time (considering that, more often, one will want to estimate the value of an observable associated with eigenmodes). Unbiased estimators of the coefficients $c_a(t)$ can be intuitively built as follows. Suppose at time $t$ we sample $N$ independent realizations of stochastic jump process whose probability distribution is described by a master equation, obtaining data $x_1,\ldots,x_N$. Since each datum was sampled with probability $p(t)$, the probability of the sample is $ p_{x_1}(t)\ldots p_{x_N}(t)$. Let $f_i = N^{-1} \sum_{n=1}^N \delta_{i,x_n}$ be the empirical distribution of such samples, that is, an histogram. Consistently, if we average the empirical distribution over possible samples we obtain the original distribution (we drop the time-dependence hereafter):
\be
\langle f_i \rangle = \sum_{x_1,\ldots,x_N} p_{x_1}\ldots p_{x_N} f_i = p_i. 
\ee
We project the empirical distribution onto the left eigenvectors $q^{L,a}$ of $W$ such that $(q^{L,a},q^b) = \delta^{ab}$. The empirical coefficients $\hat{c}_a = \sum_i q_i^{L,a} f_i$ are unbiased estimators, since one can easily show that $\langle \hat{c}_a \rangle = c_a$. Let $C_{ab}= \langle (\hat{c}_a -c _a) (\hat{c}_b - c _b)\rangle$ be their covariance matrix. The multivariate Cram\`er-Rao bound then states that $C G \geq N^{-1} I  $,
where matrix inequality $A \geq B$ means that $A-B$ is positive semidefinite. Notice that the bound becomes less strict as the number of samples is increased.

For equilibrium systems, using an orthonormal set of modes with respect to the scalar product $\langle\,\cdot\, ,\, \cdot \, \rangle$, we have $C_{aa} \geq N^{-1}$,
which is a statement about the variance of individual estimators; the information stored in the estimators ``decouples''. However, by a straightforward calculation one can show that the simple estimators that we built are not uncorrelated, even for equilibrium systems. Building uncorrelated maximum-likelihood estimators, using orthogonality of the Fisher parameters, is an important task in estimation theory \cite{miura}. In this respect the theory states that equilibrium systems are more tractable for parameter estimation.

Another case of interest is that of nearly defective systems. Using the nearly degenerate Fisher matrix in Eq.(\ref{eq:gdeg}), and evaluating $w_0^T C G w_0$ along vector $w_0 = (1,-1)$, one obtains
\be
2 c_{12} - c_{11} - c_{22} \geq \frac{2}{\epsilon^2 N},
\ee
which as $\epsilon \to 0$ implies that the covariance matrix is singular, and that the cross-correlation must diverge, as one will not be able to distinguish among the two modes. This kind of behavior has been put in contact with classical and quantum phase transitions \cite{janke,zanardi}. It is interesting to note that in the context of Markovian dynamics, this critical behavior is accompanied with polynomial terms in the time evolution \cite{polettini2}. 

\section{Conclusions and perspectives}

In this paper we showed how algebraic properties of the Fisher matrix, in a basis of decay modes, can be useful to tackle specific issues regarding nonequilibrium Markov processes, such as the monotonicity and the convexity of (candidate) Lyapunov functions.

In particular, we were able to produce counterexamples to the convexity of relative entropy with respect to the steady state in the ``nonequilibrium linear regime'', i.e. with initial conditions picked arbitrarily close to a nonequilibrium steady state. From a thermodynamic perspective, this tells us that there is no general principle of minimum nonadiabatic entropy production, which would represent the nonequilibrium analogue of a well-known stability criterion for close-to-equilibrium systems \cite{jiuli}. However, our counterexamples display a very subtle fine-tuning of the initial conditions, and for three-state systems we argued that a large separation of time scales has to be attained. If both these facts could be rigorously proven and extended to more general systems, since the nonconvex regime has the typical lifetime of the shortest decay time, such eventual transients would be proven to be completely irrelevant with respect to the dominant dynamics, and one would be able to argue that ``for all practical purposes'' such generalized principles do hold. 

Our discussion strongly relied on Fisher's information, a concept from estimation theory that has long been employed in equilibrium statistical mechanics, and is now being more and more explored with applications to nonequilibrium systems. Our use of this interesting tool was quite restrained, but we envisage that the techniques hereby introduced could be useful to discuss stability and fluctuations of nonequilibrium systems and possible relationships regarding decay modes and eigenvalues. For example, much work on the interplay between the imaginary and the real parts of complex eigenvalues has yet to be addressed. As briefly described in the last section, other interesting applications of Fisher's information might come from exploiting the full machinery of information geometry and estimation theory, in particular as regards the actual definition of unbiased estimators and the implications of the Cr\'amer-Rao bound.
Finally, normal systems might deserve further attention, as they are, in a way, the nonequilibrium analogue of detailed balance systems.

\section*{Aknowledgments}

We are grateful to Christian Maes for useful discussion, and for comments on the draft. The research was supported by the National Research Fund Luxembourg in the frame of project FNR/A11/02.
 
\appendix

\section{\label{complex}\label{normal}Theory and results: Complex spectrum}

In this section we discuss the construction of the Fisher matrix for systems with complex spectrum, describing conditions for convexity violation, and introducing a class of nonequilibrium generators (called ``normal'') for which convexity holds. They are peculiar with respect to time reversal,  in a way that might be considered a nonequilibrium generalization of detailed balance. 

We introduce the case of complex spectrum with the following chain of considerations. Let a three-state system have two real decay modes. The Fisher matrix reads
\be
G = \left(\ba{cc} \langle  {q}_+, {q}_+ \rangle & \langle  {q}_+, {q}_- \rangle \\ \langle  {q}_+, {q}_- \rangle  &  \langle  {q}_-, {q}_- \rangle\ea \right), \label{eq:2fisher}
\ee
and the square root of its determinant is the area of the parallelogram formed by the two vectors. The area of a parallelogram coincides with the area of the parallelogram formed by its diagonals, $( {q}_+ +  {q}_-)/2$ and $ {q}_- -  {q}_+$. We rescale the diagonals of factors $\sqrt{2}$ and $1/\sqrt{2}$ respectively, while keeping the area invariant, and define vectors $ {q}_1 = ( {q}_+ +  {q}_-)/\sqrt{2}$, $ {p}_2 = ( {q}_+ -  {q}_-)/\sqrt{2}$ and the tilted Fisher matrix
\be
\tilde{G} =
\left(\ba{cc} \langle  {q}_1 , {q}_1 \rangle & \langle  {q}_1, {p}_2 \rangle \\ \langle  {q}_1, {p}_2 \rangle  &  \langle  {p}_2, {p}_2 \rangle\ea \right).
\ee
We have $\det G = \det \tilde{G}$. Notice that $\tilde{G}$ is obtained from $G$ after a rotation of an angle $\pi/2$ of the defining vectors. In general, when one performs a change of basis in the space of modes, $q^a \to \sum_a A_a^b q^a$, $G$ transforms by matrix congruence $G \to A^T GA$, which is a similarity of matrices only when $A = A^{-T}$ is orthogonal. For equilibrium systems $G \propto I $; for defective systems $G$ is degenerate. Both these properties remain true for all representatives under the orthogonal transformation $A$; such properties are equivalently described by $G$ or $\tilde{G}$.

When the generator admits a couple of complex-conjugate eigenmodes
\be
 {q}_\pm =  ( {q}_1 \pm i  {q}_2)/\sqrt{2}
\ee
relative to complex-conjugate eigenvalues
\be \lambda^{\pm}  ~=~ -1/ \tau \pm i \omega,  \nonumber \ee
the matrix defined in Equation (\ref{eq:2fisher}) has complex entries. It is meaningful to  perform a rotation in the complex plane. Switching to the tilted matrix, with $q_2 = ip_2$, we have
\be
\tilde{G} =
\left(\ba{cc}1 & 0  \\ 0 &i \ea \right)
\left(\ba{cc} \langle  {q}_1 , {q}_1 \rangle & \langle  {q}_1, {q}_2 \rangle \\ \langle  {q}_1, {q}_2 \rangle  &  \langle  {q}_2, {q}_2 \rangle\ea \right)
\left(\ba{cc}1 & 0  \\ 0 &i \ea \right).
\ee
Notice that $\det \tilde{G} < 0$. A candidate as a Fisher matrix for systems with complex spectrum is then given by $G = |\tilde{G}|$. Generalizing, when the system has both real eigenvalues labelled by $a$ and complex eigenvalues labelled by $k$, we define the Fisher matrix
\be
G
~=~ \left(\ba{ccc}
\langle  {q}^{\;k}_1, {q}^{\,k'}_1\rangle  & \langle  {q}^{\;k}_1, {q}^{\,k'}_2 \rangle  & \langle  {q}^{\;k}_1, {q}^{\,a'} \rangle  \\
\langle  {q}^{\;k}_2, {q}^{\,k'}_1\rangle  & \langle  {q}^{\;k}_2, {q}^{\,k'}_2\rangle  & \langle  {q}^{\;k}_2, {q}^{\,a'} \rangle  \\
\langle  {q}^{\,a}, {q}^{\,k'}_1\rangle  &  \langle  {q}^{\,a}, {q}^{\,k'}_2\rangle & \langle  {q}^{\,a}, {q}^{\,a'} \rangle  \nonumber
 \ea \right)_{k,k',a,a'} .
\ee
Now consider the state
\be
 {p}(t) ~=~  {p}^{\,\ast} + \sum_h c_a(t)  {q}^{\,a} + \sum_k \left[ c_k(t)    {q}^{\;k}  + c^{\,\ast}_k(t)    {q}^{\,k {\,\ast}} \right]\label{eq:initial}.
\ee
Define $c_k^1 = (c_k + c_k^{\,\ast}) /\sqrt{2}$ and $c_k^2 = (c_k - c_k^{\,\ast}) /i\sqrt{2}$ and collect the data in a vector
\be
\bs{c}^T = (c_a,c^1_k,c^2_k)_{a,k}.
\ee
It is a simple exercise to prove that the relative entropy in the linear regime reads $H = \bs{c}^T G \bs{c}$.
The time-derivative of $\bs{c}$ is also easily calculated,
\be \dot{\bs{c}} =  - (\Upsilon + i\Omega)\bs{c} \label{eq:upsiev} \ee
where we introduced the matrix
\be \Upsilon + i\Omega
= - \left(\ba{cccccc} \ddots \\
& \tau_k^{-1} & \omega_k &\\
& -\omega_k & \tau_k^{-1} \\
& & & \ddots \\
& & & & \tau_a^{-1}\\
& & & & & \ddots
 \ea \right)
 \ee
and $ \Upsilon = \mathrm{diag}\, \{\tau_k^{-1},\tau_k^{-1},\tau_a^{-1}\}_{k,a}$. Exponentiating Equation (\ref{eq:upsiev}) gives the typical decaying-oscillating character. Finally, evaluating the second time derivative of the relative entropy we obtain $K= K_1+K_2$, with
\bea
K_1 & = & 2  (\Upsilon - i\Omega) G  (\Upsilon + i\Omega), \\
K_2 & = &  (\Upsilon - i\Omega)^2 G + G (\Upsilon + i\Omega)^2.
\eea

Normal systems are those whose generators commute with their time reversal:
\be
W\overline{W} = \overline{W}W. 
\ee
As already pointed out above, $W$ and $\overline{W}$ always have the same spectrum, but they might not have the same eigenvectors. Normal generators do. Let $\bar{q}^{\,k}_\pm$ be the complex conjugate eigenvectors of $\overline{W}$ wih respect to $\lambda^k_{\pm}$. Applying $W$ to the eigenequation we obtain
\be
W\overline{W}\bar{q}^{\,k}_\pm  = \overline{W}W \bar{q}^{\,k}_\pm = \lambda_{\pm}^k  \bar{q}^{\,k}_\pm, 
\ee
which implies that also $W \bar{q}^{\,k}_\pm$ is an eigenvector of $\overline{W}$, relative to eigenvalue $\lambda_{\pm}^k$. Then $\bar{q}^{\,k}_\pm$ must be an eigenvector of $W$. Now, since matrix $W\overline{W}$ must have positive spectrum (being similar to $HH^T$, which is symmetric hence with real spectrum), then one necessarily has that
\be
\bar{q}^{\,k}_\mp = q^{\,k}_\pm,
\ee
since $\lambda_{+}^k\lambda_{-}^k$  are the only real products of eigenvalues.

To resume, normal systems are such that the time reversal has the same spectrum and eigenvectors as the original dynamics, but it inverts positive and negative frequency modes. Time reversal inverts the oscillatory character (much like for quantum mechanical systems), while damping occurs in the same way.

By application of the spectral theorem to normal matrices \cite{horn}, it can be proven that for normal systems the Fisher matrix $G$ is diagonal, or in other words eigenmodes can be normalized, yielding $G = I $. Therefore we obtain
\be
K = 4 \Upsilon^2 > 0.
\ee
Hence normal systems satisfy convexity.

Moreover, for normal systems we have
\be
K_2 = 2( \Upsilon^2 - \Omega^2 ),
\ee
which is positive on its own if and only if relaxation times are smaller than the coresponding decay periods, i.e. $1/\tau_k < \omega_k^{-1}$, $\forall k$. This property seems to be valid and has already been consjecured by Maes et al. Indeed, even for non-normal systems, contrarily to the real case, we report that as we performed an intensive numerical search for three-state generators, we were not able to find systems that have a non-positive $K_2$. Hence convexity seems to be more robust for systems with complex spectrum.

\section{\label{app2}Defective three-state system}

The task is to express parameters $\alpha,\beta$ in terms of the transition rates of a generic nearly defective three-state generator
Let us introduce the quadratic oriented-spanning-tree polynomial $z$ \cite{schnak} and the linear polynomial $t$, given by minus the trace of $W$,
\bes
t & = & w_{21} + w_{31} + w_{12} + w_{32} + w_{13} + w_{23},    \\
z & = & w_{12} w_{13} + w_{32} w_{13} + w_{12} w_{23}  + w_{21} w_{13} + w_{21} w_{23} \nonumber  \\
& & + \, w_{31}w_{23} + w_{31} w_{12} + w_{21} w_{32} + w_{31} w_{32}.  
\ees
The system has a degenerate spectrum when $t^2 = 4z$. We perturb to first order the eigenvalues near the degenerate spectrum, $\lambda_{\pm} = \tfrac{1}{2} \left( -t \pm \sqrt{t^2 - 4 z} \right) \approx - \sqrt{z} (1 \mp \epsilon)$, from which $\lambda = -\sqrt{z}$.
The steady state and the decay modes are given by
\bea
{p}^\ast & = & \frac{1}{z}\left( \ba{c} w_{12} w_{13} + w_{32} w_{13} + w_{12} w_{23}  \\ w_{21} w_{13} + w_{21} w_{23} + w_{31}w_{23} \\ w_{31} w_{12} + w_{21} w_{32} + w_{31} w_{32} \ea \right),\\
q_{\pm} & = & \left( \ba{c} (w_{13}+w_{23}+\lambda_{\pm})(w_{12}+w_{32}+\lambda_{\pm}) - w_{23} w_{32}  \\ w_{23}w_{31}  + w_{21} (w_{13} + w_{23} + \lambda_{\pm}) \\ w_{32} w_{21} + w_{31} (w_{12} + w_{32} + \lambda_{\pm}) \ea \right),  \nonumber
\eea
wherefrom one can read off values of $x,y$ in terms of the transition rates. A more compact representation can be given as follows. Letting $e_1 = (1,0,0)$, and ${w}_1$ be the first column of the generator, we can express the decay modes as ${q}_{\pm} =  {p} ^{\, \ast} - e_1 + z^{-1} \lambda_{\pm}  {w}_1$, 
which can be proven by pugging this expression into the eigenvector equation, and multiplying by $\lambda_{\mp}$,
\be
\lambda_{\mp} (W  - \lambda_{\pm}) ( p^{ss} - \hat{e}_1 + z^{-1} \lambda_{\pm} {w}_1 ) =  (W +  t )  {w}_1 + z (\hat{e}_1 - {p}^{\, \ast} ),
\ee
where we used $\lambda_+ \lambda_- = z$ and $\lambda_+ + \lambda_- = -t$. The above expression can be shown to vanish by direct calculation. We then obtain
\be
x = {p} ^{\, \ast} - \hat{e}_1 - {w}_1/\sqrt{z}, \qquad y = {w}_1/ \sqrt{z}, 
\ee
and similarly we can express parameters $\alpha,\beta$ in terms of the transition rates.

\end{document}